# Drastic Improvement of Air Stability in an n-type Doped Naphtalene-diimide Polymer by Thionation

*Diego Nava[†§], Younghun Shin[‡], Matteo Massetti[†§], Xuechen Jiao[¥¤], Till Biskup[#], Madan S. Jagadeesh[§], Alberto Calloni[§], Lamberto Duò[§], Guglielmo Lanzani[†§], Christopher R. McNeill[¥], Michael Sommer[‡\*], Mario Caironi[†\*]*

† Center for Nano Science and Technology @PoliMi, Istituto Italiano di Tecnologia, Via Pascoli 70/3, Milano 20133, Italy.

§ Politecnico di Milano, Dipartimento di Fisica, P.za L. da Vinci 32, Milano 20133, Italy

‡ Institut für Chemie, Technische Universität Chemnitz, Straße der Nationen 62, 09111 Chemnitz, Germany

¥ Department of Materials Science and Engineering, Monash University, Wellington Road, Clayton, VIC, 3800, Australia

¤ Australian Synchrotron, 800 Blackburn Road, Clayton, VIC, 3168, Australia

# Institut für Physikalische Chemie, Albert-Ludwigs-Universität Freiburg, Albertstraße 21, 79104 Freiburg, Germany

\* Mario Caironi: mario.caironi@iit.it

\* Michael Sommer: michael.sommer@chemie.tu-chemnitz.de

**Abstract**

Organic thermoelectrics are attractive for the fabrication of flexible and cost-effective thermoelectric generators (TEGs) for waste heat recovery, in particular by exploiting large-area printing of polymer conductors. Efficient TEGs require both p- and n-type conductors: so far, the air instability of polymer n-type conductors, which typically loose orders of magnitude in electrical conductivity ($\sigma$) even for short exposure time to air, has impeded processing under ambient conditions. Here we tackle this problem in a relevant class of electron transporting, naphthalene-diimide co-polymers, by substituting the imide oxygen with sulphur. n-type doping of the thionated co-polymer gives rise to a higher $\sigma$ with respect to the non-thionated one, and most importantly, owing to a reduced energy level of the lowest-unoccupied molecular orbital, $\sigma$ is substantially stable over 16 h of air exposure. This result highlights the effectiveness of chemical tuning to improve air-stability of n-type solution-processable polymer conductors and shows a path towards ambient large-area manufacturing of efficient polymer TEGs.

**Introduction**

Energy harvesting with organics sees its most mature example in organic solar cells, which enables lightweight and large-area solar energy converters suitable for distributed energy generation.[1,2] Organic materials could also support the development of other energy harvesting and scavenging[3] devices that necessitate cost-competitive technologies to be economically viable, such as thermoelectric generators(OTEGs), that could be employed to supply power to low-consumption distributed and/or portable electronic devices At the same time, such development could lead to custom-shaped active coolers,[4,5] facilitating their integration into existing electronic appliances. One of the most appealing paths towards such applications is to make use of solution-processed organic compounds in order to enable mass-printed OTEGs, thus drastically limiting production costs. Yet, the development of organic thermoelectrics is far less mature than photovoltaics, with critical shortfalls in fundamental knowledge regarding the thermoelectric properties of conjugated molecule-based films,[6,7] available materials,[8–10] processing[11–13] and device engineering.[14–18]

The efficiency of the conversion of a heat flux into a current by thermoelectric (TE) materials can be related to the dimensionless material figure of merit *zT*, defined as:

(1) $\quad zT = \frac{S^2 \sigma}{\kappa} T$

where σ is the electrical conductivity, *S* is the Seebeck coefficient and κ is the thermal conductivity.

One of the most attractive characteristics of organic conductors or doped organic semiconductors is their low *κ*, typically below 1 Wm$^{-1}$ K$^{-1}$,[9,19] as a result of the suppression of the phonon component in the thermal conductivity due to the intrinsic structural disorder.[20] A common strategy therefore to improve TE properties of organic semiconductors is to improve the numerator of *zT*, namely the power factor (PF).

(2) $\quad PF = S^2\sigma \ \ [\text{Wm}^{-1}\,\text{K}^{-2}]$

Since in their pristine form conjugated organic materials are typically insulators or semiconductors with low background conductivity, doping is necessary to achieve suitable electrical conductivity.[21] Thus, tuning the doping level, namely modulating the charge carrier density, is a key aspect for optimizing $PF$ and $zT$ values.[22] In organic semiconductors the doping process relies on the addition of sub-stoichiometric amounts of a redox-active species to the host semiconductor matrix. The dopant, which is a molecular species or a salt,[23,24] donates an electron to the host (n-doping) or accepts an electron from the host, leaving a hole behind (p-doping) as a consequence either of direct charge-transfer,[25–28] or of an indirect process. For example, proton transfer from acids or hydride transfer can lead to p-type and n-type doping, respectively.[29]

For the realization of an efficient thermoelectric device, complementary p-type and n-type conducting materials with high $PF$ are needed at the same time.[30] While solution processable organic p-type materials having a $PF$ in excess of 100 $\mu\text{Wm}^{-1}\,\text{K}^{-2}$ have been demonstrated,[31–34] the major limitations are ascribable to n-type materials, for which examples combining high electrical conductivity and solution processability are very scarce. A further strong limitation for n-type materials arises from their air instability, which precludes ambient processing of OTEGs and impose severe constraints regarding devices encapsulation.[35]

To overcome this issue, both the dopant and the semiconductor must be air-stable. With regard to the dopant, adopting species with very low ionization energy to induce an electron transfer to the host semiconductor is critical, because the donating molecule is very prone to redox reactions with species in the atmosphere.[36] This is the reason why the hydride transfer scheme proposed by Bao and co-workers from ambient stable benzimidazole derivatives, such as 4-(1,3-dimethyl-2,3-dihydro-1H-benzoimidazol-2-yl)-N,N-dimethylaniline (N-DMBI) and 4-(1,3-Dimethyl-2,3-

dihydro-1H-benzoimidazol-2-yl)-N,N-diphenylaniline (N-DPBI), immediately gained interest.[37] Chabinyc and co-workers reported a maximum conductivity of σ $10^{-3}$ Scm$^{-1}$ for poly{[N,N'-bis(2-octyldodecyl)-naphthalene-1,4,5,8-bis(dicarboximide)-2,6-diyl]-alt-5,5'-(2,2'-bithiophene)}, here abbreviated as PNDIT2 (Figure 1a), doped with N-DMBI.[38] Koster and co-workers[39,40] recently explored a benzimidazole derivative to dope a modified fullerene, both tailored with hydrophilic triethylene glycol type side chains, achieving an electrical conductivity of 2.05 Scm$^{-1}$ and a *PF* of 19.1 μW m$^{-1}$ K$^{-2}$, one of the highest reported for solution processable n-type materials. Pei and co-workers[41] reported a series of BDPPV n-type conjugated polymers with low lying LUMO levels and very high *PF* values up to 28 μW m$^{-1}$ K$^{-2}$. However, all the previous systems did not exhibit air stability or at least this aspect was not investigated. Generally, it is assumed that ambient stability of n-type conductivity can be achieved using materials with a LUMO energy level more negative than −3.9 eV.[42–46]

The lowering of the LUMO energy of the semiconductor can be obtained by diverse modifications of the electron withdrawing cores such as in naphthalene diimide (NDI) based co-polymers, one of the most studied class of n-type materials [47–55] With the core positions being partially blocked in PNDIT2 by the comonomer, the strategy of lowering the LUMO by replacing the imide group by a mono-thioimide group, known as thionation, appears highly suitable for the development of stable n-doped small molecule NDIs [56–59] and NDI-based polymers.[60,61]

Here we make use of thionated PNDIT2, referred to as 2S-*trans*-PNDIT2 (Figure 1b) in which two carbonyl oxygens are substituted by sulphur with full *trans*-regioselectivity.[61] 2S-*trans*-PNDIT2 exhibits a lower LUMO level compared to PNDIT2 which enables a drastically improved stability of n-type electrical conductivity at ambient conditions. When doped with N-DPBI (Figure 1c), σ is as high as 6·10$^{-3}$ Scm$^{-1}$, with an improved power factor with respect to PNDIT2, reaching $4.9 \cdot 10^{-2}$ µWm$^{-1}$K$^{-2}$. Importantly, while doped PNDIT2 films lose several orders of magnitude of $\sigma$ within 200 minutes, reaching almost the low conductivity of the undoped film, the conductivity of 2S-*trans*-PNDIT2 decreases by only a factor of two after 16 hours of continuous air exposure.

**Materials**

The herein used 2S-*trans*-PNDIT2 was obtained from the educt PNDIT2 using Lawesson's reagent as previously published.[61] This PNDIT2 has a molecular weight from size exclusion chromatography (SEC) $M_{n,SEC}/M_{w,SEC}$ = 12/19 kg/mol, and was also used as reference herein. The molecular weight of 2S-*trans*-PNDIT2 obtained by SEC under the same conditions was $M_{n,SEC}/M_{w,SEC}$ = 8/355 kg/mol. However, for an equal chain length of the two materials it is apparent that the molar mass of 2S-*trans*-PNDI must be slightly higher than that of the corresponding educt PNDIT2. In fact molar mass determination by SEC is not straightforward here, as also seen by the very high $M_w$ values caused by aggregation. We therefore made additional use of NMR end group analysis, which gives a $DP_{n,NMR}$ of 8.7 for PNDIT2 (and the same value for 2S-*trans*-PNDIT2).[61] From the masses of the repeat units and an O/S conversion of 96 % for 2S-*trans*-PNDIT2, we estimate absolute number average molecular weights $M_{n,NMR}$ of 8.6 and 8.9 kg/mol for PNDIT2 and 2S-*trans*-PNDIT2, respectively.

**Optical and Electrical Properties**

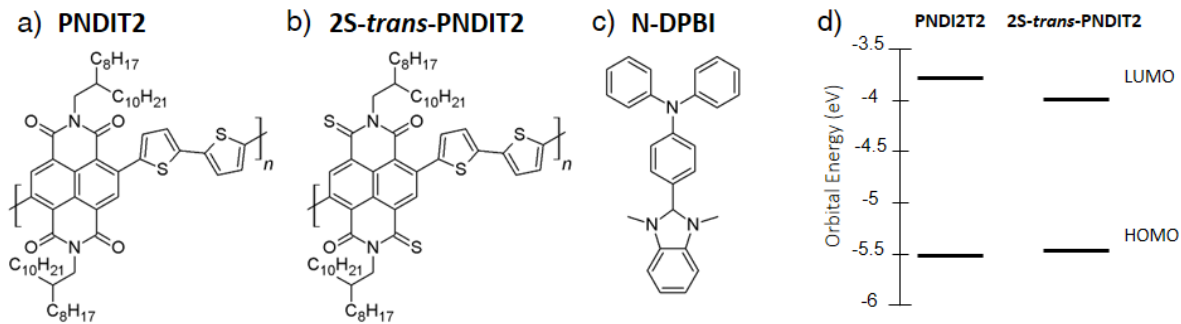

**Figure 1.** Chemical structures of a) PNDIT2, b) the thionated derivative 2S-*trans*-PNDIT2, c) the hydride dopant N-DPBI. d) Energy diagram for frontier energy orbitals of both co-polymers obtained from cyclic voltammetry and absorption in solution.

The effect of the substitution of the imide oxygen atoms with sulphur atoms is clearly visible from the UV-visible absorption spectra (see Supporting Information, SI, Figure S1)[61]. Comparing the spectra of the PNDIT2 and 2S-*trans*-PNDI, a strong red-shift of ≈100 nm in the absorption on-set is observed for the latter, indicating a significant reduction of the optical energy gap. Since the HOMO is mostly located on the thiophene moieties,[62] while the LUMO is localized on the naphthalene-diimide unit,[63] the energy gap reduction should be ascribable to a deeper LUMO energy level as an effect of thionation. This is confirmed by cyclic voltammetry measurements (Figure S2), as reported by Shin *et. al.*[61] By combining optical data and electrochemical data, following the methodology described in the SI, we can obtain a first estimation of the HOMO/LUMO energies (Figure 1d). For 2S-*trans*-PNDIT2 we extract energy values of -5.47 eV for the HOMO level and -3.96 eV for the LUMO, while for PNDIT2 we obtain -5.51 eV and -3.75 eV respectively for the HOMO and the LUMO level. As expected, thionation of the acceptor moiety mostly reduces the LUMO level of 2S-*trans*-PNDIT2, while leaving the HOMO level substantially unperturbed.

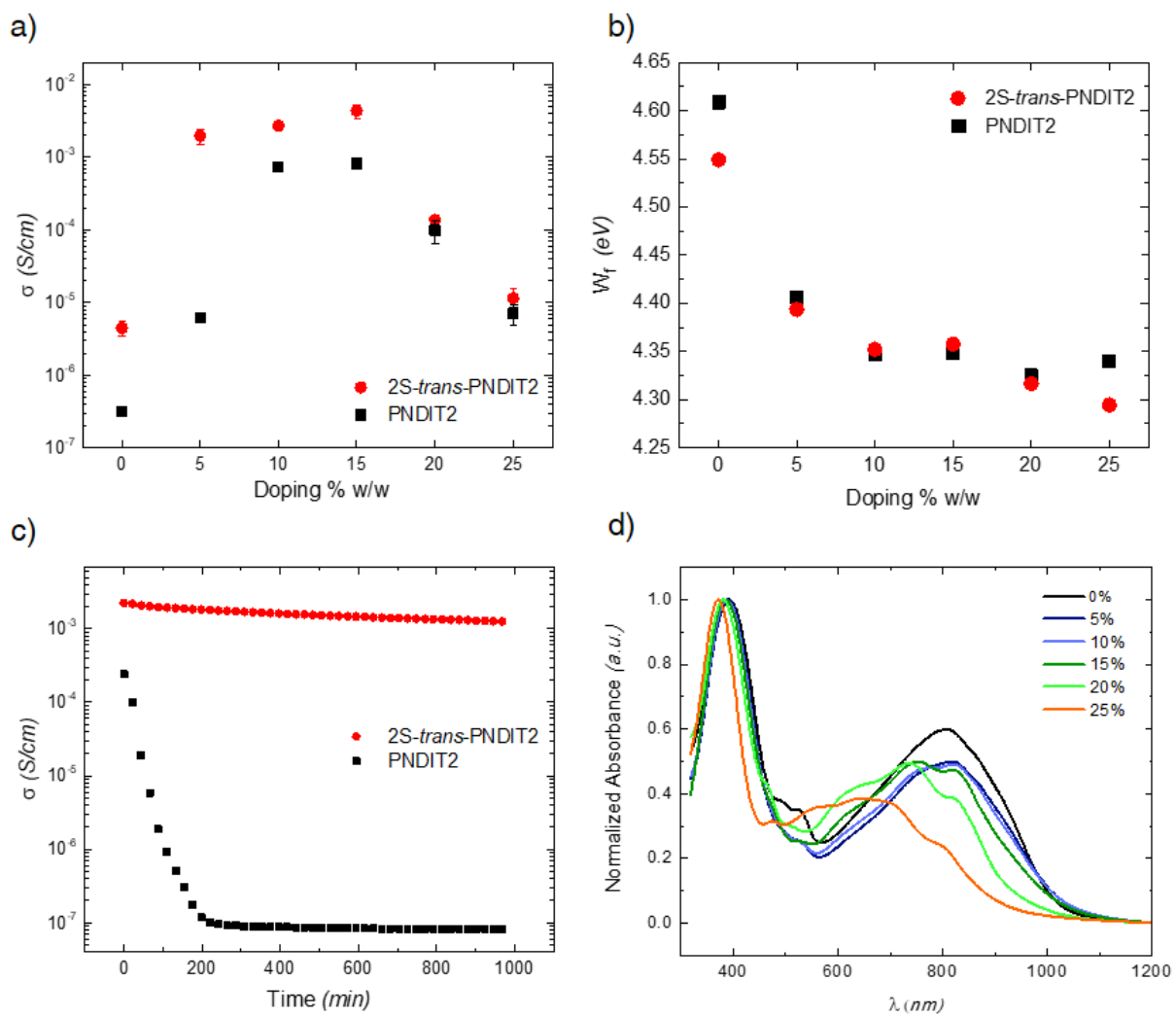

**Figure 2.** a) Electrical conductivity of 2S-*trans*-PNDIT2 (red dots) and PNDIT2 (black squares) thin films as a function of the N-DPBI dopant concentration. Each data point is an average of at least four devices, vertical error bars represent the standard deviation. b) Work function plot for 2S-*trans*-PNDIT2 (dots) and PNDIT2 (squares) as a function of doping concentration. c) Time dependence of the electrical conductivity of non-encapsulated thin films of 2S-*trans*-PNDIT2 (red

dots) and PNDIT2 (black squares), with 15 % w/w N-DPBI, upon ambient air exposure. d) UV-Vis-NIR thin film absorption spectra of 2S-*trans*-PNDIT2 doped with different w/w concentration of N-DPBI.

In order to dope the polymer films, we selected N-DPBI. Following the procedure reported by Schlitz *et al.* for PNDIT2,[38] we added N-DPBI directly to the 2S-*trans*-PNDIT2 toluene solution at various concentrations, from 5 % to 25 % by weight, and deposited a 40 nm thick film by spin-coating. We applied the same doping procedure to PNDIT2 as a reference.

In its pristine, undoped state, the 2S-*trans*-PNDIT2 film shows an electrical conductivity of $(4\pm2)\cdot10^{-6}$ Scm$^{-1}$. For undoped PNDIT2, a value that is one order of magnitude lower is found $(3\pm1)\cdot10^{-7}$ Scm$^{-1}$. This is in agreement with electron spin resonance (ESR) spectroscopy data indicating the presence of unpaired electrons in 2S-*trans*-PNDIT2, but not in PNDIT2 (Figure S11). After the addition of only 5 % w/w of N-DPBI, the conductivity of 2S-*trans*-PNDIT2 films drastically increases by more than two orders of magnitude, reaching $(2.0\pm0.1)\cdot10^{-3}$ Scm$^{-1}$. An increase of the doping concentration above 5 % w/w, at first leads to a small increase of $\sigma$, reaching a maximum value of $(5.9\pm0.2)\cdot10^{-3}$ Scm$^{-1}$ at 15 % w/w of N-DPBI. A further increase in dopant concentration is then detrimental for the electrical conductivity. A similar dependence of $\sigma$ on dopant concentration was observed for PNDIT2 reference films, in good agreement with literature,[38,64–66] where the maximum $\sigma$ of $(1.2\pm0.2)\cdot10^{-3}$ Scm$^{-1}$, achieved at 15% w/w, is five times lower than for 2S-*trans*-PNDIT2. The difference in $\sigma$ between the two systems is much stronger at low doping concentration: at 5 % w/w, $\sigma$ is 330 times higher in the thionated sample than in the parent co-polymer, and even two times higher than the maximum electrical conductivity achieved at 15 % w/w in PNDIT2.

The effect of doping was also monitored by measuring the work function ($W_f$) by the Kelvin Probe technique (Figure 2b).). Undoped films of 2S-*trans*-PNDIT2 feature a lower $W_f$, in agreement with an increased number of excess electrons already seen with ESR. With increasing dopant concentration, $W_f$ reduces from 4.55 eV in the pristine film to 4.35 eV in the doped film at 10 % w/w concentration, where it reaches a plateau; a further increase in the doping concentration does not lead to an appreciable reduction of the work function as detectable by our Kelvin Probe setup. X-ray photoemission spectroscopy (XPS) was used to certify the presence of dopant molecules and to assess their chemical state. In particular, in the case of doped 2S-*trans*-PNDIT2 films, we detected the spectroscopic features of oxidized N-DPBI molecules (Figure S4 in SI and details therein).

Besides the difference in $\sigma$, what is remarkable is the difference in stability of the electrical conductivity between the two systems when films are directly exposed to ambient air. Figure 2c reports $\sigma$ as a function of the ambient air exposure time, at room temperature and relative humidity of about 50 %, for both materials at a doping concentration of 15 % w/w. The electrical conductivity of PNDIT2 rapidly drops, reaching almost the pristine film value after 200 minutes. Such strong instability basically prevents any ambient processing of this doped system and its use for OTEGs fabrication. On the contrary, the electrical conductivity of 2S-*trans*-PNDIT2 remains within the same order of magnitude of its initial value for as long as 16 hours of air exposure, showing only a reduction factor slightly lower than 2. Similar behaviour in air stability was observed in 2S-*trans*-PNDIT2 for 5% w/w and 10% w/w doping concentrations (Figure S10). We assign the enhanced stability to the lower LUMO level in 2S-*trans*-PNDIT2 that allows electrons to relax at energy levels less prone to redox processes in ambient air. This is an unprecedented stability for doped n-type organic materials, and would clearly leave room for ambient processing of organic thermoelectric devices, strongly desirable in case of high-throughput printing processes, before encapsulation.[46]

It is interesting to note that the field-effect mobility as measured in bottom-contact, top-gate field-effect transistors based on the pristine thin films, is higher for PNDIT2 than for 2S-*trans*-PNDIT2, reaching the values of 0.30±0.02 cm$^2$V$^{-1}$s$^{-1}$ and 0.05±0.01 cm$^2$V$^{-1}$s$^{-1}$, respectively (details reported in Figure S5).[61] The bulk mobility of doped films and the mobility of field-effect transistors are not directly comparable.[67,68] However, since the charge density achieved in the field-effect device is of the same order of magnitude of the one achieved in doped films,[69,70] it is reasonable to expect that the higher conductivity in doped 2S-*trans*-PNDIT2 films (Figure 3a) is not caused by an improved electron mobility.

Thus, the improved $\sigma$ in doped 2S-*trans*-PNDIT2 films suggests a higher efficacy of the doping process leading to a higher charge density. The ESR signal intensity for 2S-*trans*-PNDIT2 at 15% w/w doping was two times higher than the one of PNDIT2, clearly confirming a higher carrier density (Figure S12).

To further investigate the factors that cause the higher conductivity of 2S-*trans*-PNDIT2, we followed the doping process through optical absorption and structural measurements on thin films as a function of doping concentration. Figure 2d shows the UV–vis optical absorption spectra evolution of 2S-*trans*-PNDIT2 with doping concentration. The pristine film has a broad band from 600 to 1000 nm, with a peak around 810 nm. The strong red-shift with respect to the spectrum of molecularly dissolved chains in chloronaphthalene solution is due to aggregation.[62] At low doping concentration (5 and 10 % w/w), the intensity of the low energy band reduces, with the appearance of a stronger component at around 600 nm in the 10 % w/w case, indicating an increased presence of a non-aggregated, amorphous phase. For a doping concentration above 15 % w/w, the shoulder between 800 nm and 1000 nm becomes increasingly suppressed in favour of the non-aggregated component at 600 nm and the ratio between the 750 nm and 830 nm peaks rapidly increases. Overall, the evolution of the UV-Vis spectra with doping concentration suggests that the dopant molecule suppresses chain-chain interactions of 2S-*trans*-PNDIT2, resulting in an increased amorphous component. Differently, in case of doped PNDIT2 (Figure S3), an increasing doping concentration mainly causes a reduction of the intensity of the 600 – 800 nm band while keeping the ratio of vibronic bands constant up to 15 % w/w. Higher doping concentrations did not lead to further modification of the UV-vis spectra indicating a weaker interaction of the dopant with PNDIT2.

Optical absorption measurements therefore suggest a structural evolution in 2S-*trans*-PNDIT2 films with doping.

**Morphological Characterization**

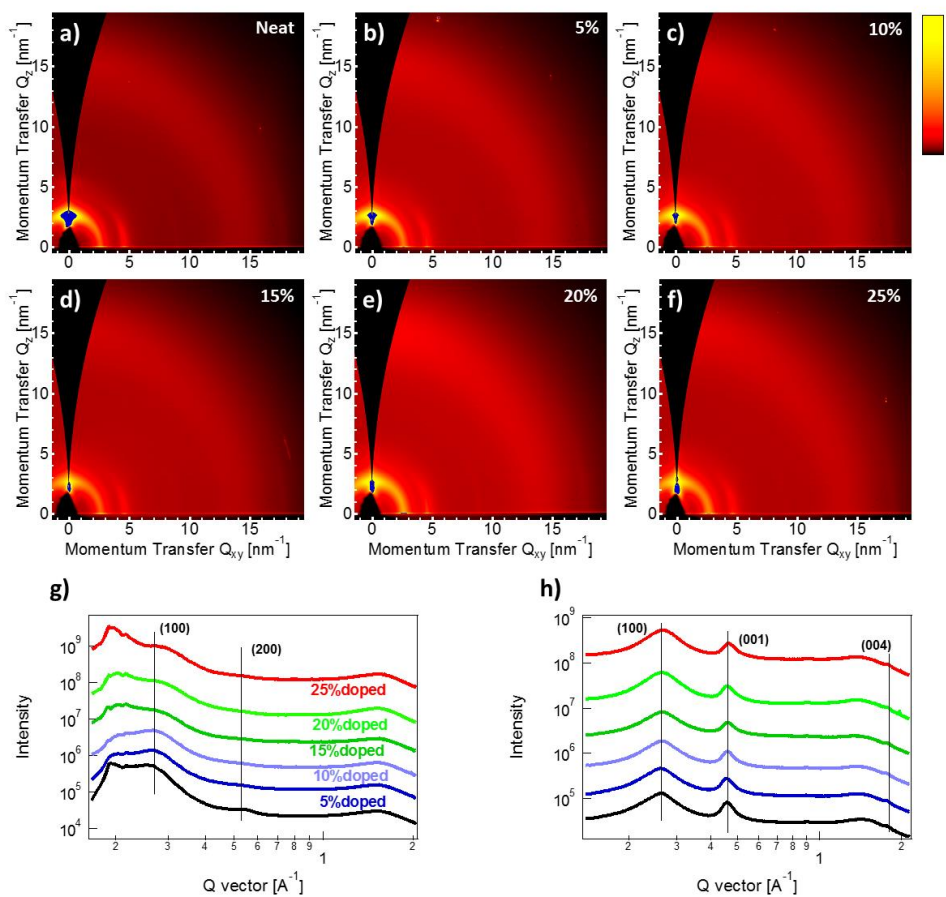

**Figure 3.** Two-dimensional data plots of GIWAXS measurements of 2S-*trans*-PNDIT2 films for different doping concentrations, from a) to f): 0, 5, 10, 15, 20 and 25 % w/w. Corresponding one-dimensional g) out-of-plane and h) in-plane GIWAXS profiles.

The microstructure of these thin films was further investigated by grazing incidence wide angle X-ray scattering (GIWAXS). Comparing the GIWAXS patterns of films of undoped 2S-*trans*-PNDIT2 (Figure 3a) and PNDIT2 (Figure S4a), 2S-*trans*-PNDIT2 exhibits less microstructural order. Although lamellar packing and backbone repeat distance are similar, 2S-*trans*-PNDIT2 exhibits a lower degree of crystalline order evidenced by broader peaks with an apparent absence of a π-π stacking peak.[61] Furthermore, films of 2S-*trans*-PNDIT2 exhibit more orientational disorder, with crystallites exhibiting a broad orientational distributions from face-on to edge-on. Thus, thionation of PNDIT2 reduces microstructural order which may explain in part the reduced field-effect mobilites of 2S-*trans*-PNDIT2 compared to PNDIT2.[61] With increasing doping concentration, 2S-*trans*-PNDIT2 displays a further decrease in order. Undoped films (Figure 3a), while exhibiting a range of molecular orientations, show a slight preference for edge-on orientation, evidenced by the first-order lamellar stacking peak, (100), at ~ 2.5 nm$^{-1}$ along the out-of-plane direction, with a second-order peak, (200), also observable at ~ 5 nm$^{-1}$ (see also in-plane traces in Figure 3g). With increased doping, this edge-on population is less evident, with no preference for face-on or edge-on orientation in films with 15% w/w doping and above. The lamellar stacking d-spacings for the remaining crystalline population appear unaffected (Figure 3h), indicating that the dopant molecule is not intercalating within the side-chains. Since a π-π stacking peak is not clearly present, it is not possible to say whether the dopant molecules are intercalating between two stacked backbones. A strikingly different behaviour with doping is observed for the parent PNDIT2 (Figure S6). As-cast PNDIT2 films exhibit a pronounced edge-on stacking of chains, with four orders of lamellar stacking peaks evident. We note that this is unusual for PNDIT2 films, likely due to the low molecular weight adopted in this case. With doping, the PNDIT2 appears to become more ordered, evidenced by narrower lamellar stacking and backbone peaks, unlike for 2S-*trans*-PNDIT2, where the doping induces more structural disorder. Similar to 2S-*trans*-PNDIT2, the observed d-spacings are unaffected by doping, indicating that the dopant is not intercalating within the crystals. In both cases no scattering from the dopant crystals is appearing, with the second-order lamellar stacking no longer evident.

Taking together the results from optical absorption and GIWAXS, we conclude that N-DPBI dopant is predominantly incorporated in the amorphous portion of 2S-*trans*-PNDIT2 films and does not substantially alter the molecular packing motif of the crystals that do form. Since 2S-*trans*-PNDIT2 features less microstructural order, miscibility with N-DPBI is likely increased. Addition of N-DPBI thus hinders the crystallisation of 2S-*trans*-PNDIT2, in a similar fashion to the way fullerene molecules such as PCBM hinders the crystallisation of poly(3-hexylthiophene) in as-cast P3HT/PCBM blends.[71] A higher 2S-*trans*-PNDIT2 amorphous fraction in blends with N-DPBI is consistent with the appearance of a non-aggregated absorption component in UV-vis spectra with increasing dopant concentration. A reduction in the extension of polymer crystallites upon dopant addition results in a smaller fraction of the film which is precluded to the dopant and can explain the higher conductivity at 5 % w/w doping with respect to PNDIT2, despite a lower field-effect mobility. Still, at high dopant concentration, miscibility of the two components is a limiting factor on the doping efficacy, with an obvious phase separation, also evidenced by AFM topography measurements (Figure S7).[38] For PNDIT2 films, the high tendency of PNDIT2 chains to crystallise means that there is a lower fraction of chains which are able to interact with the dopant on the molecular level.

**Seebeck Measurements**

The Seebeck coefficient of the two doped co-polymers was finally measured at 15% w/w. For 2S-*trans*-PNDIT2 thin films doped with N-DPBI at 15 % w/w, we obtained $S = -90\pm4$ µVK$^{-1}$, leading to a power factor of $4.9 \cdot 10^{-2}$ µWm$^{-1}$K$^{-2}$. The negative sign of the Seebeck is indicative of electrons being the majority carriers. *PF* is significantly higher than the maximum value we can obtain with PNDIT2, where we extract PF = $1.2 \cdot 10^{-2}$ µWm$^{-1}$K$^{-2}$ under the same conditions, in line with the results reported by Wang et al.[64] Thus, thionation of PNDIT2 is not only effective in drastically improving the air stability of doped NDI-based copolymer films but also in improving the overall thermoelectric performance. The improved electrical conductivity, owing to a more favourable interaction of the dopant and the polymer, enables a four-fold improvement in power factor to be realised.

**Conclusions**

In conclusion, we have reported an effective strategy to strongly improve ambient stability of n-type doped NDI-based copolymers. The lower LUMO level of 2S-*trans*-PNDIT2 reduces the reactivity of excess electrons in air significantly. 2S-*trans*-PNDIT2 also has a weaker tendency to crystallise compared to PNDIT2, which promotes favourable molecular interactions between the polymer backbone and the dopant. Doping of 2S-*trans*-PNDIT2 with the stable N-DPBI molecule produces a strong increase of electrical conductivity already at a doping concentration as low as 5 % w/w, and the *PF* extracted at maximum $\sigma$ is four times higher than PNDIT2. The stability of $\sigma$ for doped 2S-*trans*-PNDIT2 films completely exposed to air is unprecedented for n-type conductivity in polymers. 2S-*trans*-PNDIT2 shows a reduction in conductivity of less than a factor of two over 16 h of continuous air exposure, compared to the three orders of magnitude loss measured in the first hour only for PNDIT2. Such a pronounced improvement demonstrates that inherent environmental stability of n-type doped polymer conductors can be largely controlled through the proper design of the conjugated semiconductor system. This result is an important step towards stable n-type doped polymers, which more favourably allow ink formulation for large-area manufacturing through scalable printing technologies, which finally may enable high throughput fabrication of low-cost and efficient organic thermoelectric generators under ambient conditions based on thermocouples combining n-type and p-type materials.

**Experimental Section**

*Materials*

PNDIT2 and 2S-*trans*-PNDIT2 were synthesized according to published procedures.[61,62] N-DPBI 4-(1,3-Dimethyl-2,3-dihydro-1H-benzoimidazol-2-yl)-N,N-diphenylaniline was purchased from Sigma-Aldrich and used as received.

*Solutions preparation*

Both copolymers were dissolved in toluene at a concentration of 5 g/l and stirred at 80 °C for 4 hours. N-DPBI solutions were prepared at concentrations of 3, 5 and 8 g/l and used after ≈ 12 h of dissolution at room temperature. Aliquots of polymer and dopant solution were mixed and stirred for 10 min. at room temperature just before the deposition.

*Thin films preparation*

Low alkali 1737F Corning glass was used as substrate. The substrates were cleaned in ultrasonic bath of Milli-Q water, acetone and isopropyl alcohol, 10 minutes for each step, and exposed to $O_2$-plasma at 100 W for 10 minutes. Electrodes were obtained by thermal evaporation, through a shadow mask, of a 1.5nm thick Cr adhesion layer and 25 nm thick Au film. For Field Effect Transistor bottom electrodes were patterned by a lift-off photolithographic process, defining a 20 μm channel length and a 2 mm channel width. Metals were deposited by thermal evaporation: 1.5 nm thick Cr, as adhesion layer and 20 nm thick Au. Thin polymer films were spin cast from solutions onto pre-patterned substrates, in a nitrogen-filled glovebox at 1000 rpm for 60 s, and annealed at 150 °C for 2 hours in inert atmosphere. For Field Effect Transistor devices after the polymer thin film deposition, a dielectric layer with thickness of 600 nm was obtained by spin-coating a 80 g/l PMMA solution in n-Butyl acetate at 1300 rpm for 60 s under nitrogen atmosphere, followed by an annealing at 80 °C for 1 h in inert atmosphere. The 40 nm thick gate contact was obtained by thermal evaporation of Al through a shadow mask.

*Electrical characterization*

The I-V curves and the FET characteristics were measured at room temperature in a $N_2$ atmosphere, or in ambient air, Figure 2c, on a Wentworth Laboratories probe station with a semiconductor device analyser (Agilent B1500A). The electrical conductivity was extracted through the linear fit of the I-V characteristics, Figure S8. The saturation mobility values were extracted using the gradual-channel approximation.

*UV-Vis-NIR measurements*

UV-Vis-NIR measurements were carried out on a Perkin Elmer λ1050 spectrophotometer, using a tungsten lamp as source.

*Kelvin Probe measurements*

Kelvin Probe measurements were done in air with a KP Technology Ambient air Kevin Probe instrument. For the extraction of the work function we used the multiple single-point measurements mode. The work function of the sample was obtained by adding to the measured work function value the Contact Potential Difference (CDP) of the tip, determined with an initial calibration on standard gold substrate.

*ESR measurements*

Films were prepared under inert atmosphere by drop-casting 20 μL solution (concentration of 5 g/l) onto synthetic quartz glass substrates (Ilmasil PS, QSIL GmbH) with dimensions of 3×25 mm, followed by annealing at 150 °C for 120 min after films were dried completely. The annealed films were placed into synthetic quartz glass tubes (Ilmasil PS, QSIL GmbH) with 3.8 mm outer and 3.0 mm inner diameter and the tubes sealed afterwards. ESR spectra were recorded at room temperature on an Elexsys 580 (Bruker Biospin GmbH) spectrometer equipped with a 4119HS-W1 (Bruker) cavity. Microwave frequency: 9.800 GHz; microwave power: 150 μW (30 dB attenuation, 150 mW source power); modulation frequency: 100 MHz; modulation amplitude: 0.1 mT.

*GIWAXS Characterization*

Grazing-incidence wide-angle X-ray scattering (GIWAXS) was performed at the SAXS/WAXS beamline at the Australian Synchrotron. [72] 11 keV photons were used with scattering patterns recorded on a Dectris Pilatus 1M detector. Images shown were acquired at an incident angle close the critical angle. Such images were chosen from a series of recorded 2D patterns with incident X-ray angle varying from 0.05° to 0.25° in the step of 0.01°. The X-ray exposure time was 1 s such that no film damage was identified. The sample-to-detector distance was calibrated using a silver behenate sample. The results were analysed by an altered version of the NIKA 2D [73] based in IgorPro.

*Atomic Force Microscopy*

AFM polymer thin film samples were prepared using the same procedure describe in the experimental section. The surface morphology of the films was obtained with an Agilent 5500 Atomic Force Microscope operated in the Acoustic Mode.

*Seebeck measurements*

For the Seebeck measurements we employed a home-made set up described by Beretta et al.[74] The measurements were done in vacuum, at $10^{-4}$ mbar, to avoid convection phenomena and to preserve the polymer electrical properties. In addition, the measurements were performed at a temperature of 45 °C, to have an electrical resistance below 10 MΩ, which is our system threshold for a correct measurement of the Seebeck coefficient. Before the measurement, we annealed overnight the samples at 90 °C, to avoid any possible thermal effects and/or thermal hysteresis during the thermoelectric measurements.

**Supporting Information Available:**

Optical and electrochemical characterization of both co-polymers, extraction procedure of HOMO and LUMO levels, XPS analysis, Field Effect Transistors characterization, GIWAXS characterization of PNDIT2, AFM images and Current/Voltage plots, ESR spectra.

**Notes**

The authors declare no competing financial interest.

**Acknowledgements**

The authors thanks F. Scuratti for help with AFM measurements. M. C. and D. N. acknowledge the European Research Council (ERC) for the financial support, under the European Union's Horizon 2020 research and innovation programme 'HEROIC', grant agreement 638059. Y. S. and M. S. acknowledge the Deutsche Forschungsgemeinschaft (SO 1213/ 8-1) for funding. T.B. acknowledges the Deutsche Forschungsgemeinschaft (TB 1249/3-1) for funding. This work was performed in part at the SAXS/WAXS beamline at the Australian Synchrotron, part of ANSTO. C.R.M. acknowledges support from the Australian Research Council (DP170102145).

(74) Beretta, D.; Massetti, M.; Lanzani, G.; Caironi, M. Thermoelectric Characterization of Flexible Micro-Thermoelectric Generators. *Rev. Sci. Instrum.* **2017**, *88*, 015103.

*TOC Graphic*

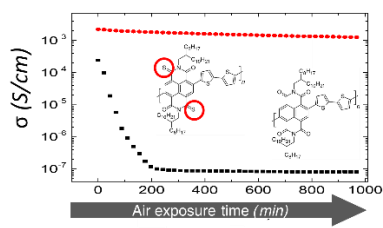